%%%%%%%%%%%%%%%%%%%%%%%%%%%%%%%%%%%%%%%%%%%%%%%%%%%%%%%%%%%%%%%%%%%%%%%%%%
%%%      This is a plain tex file, which using (input) phyzzx.tex      %%%
%%%  and phyzzx.fon, and it can be compiled in "Win Edit" running Tex. %%%
%%%%%%%%%%%%%%%%%%%%%%%%%%%%%%%%%%%%%%%%%%%%%%%%%%%%%%%%%%%%%%%%%%%%%%%%%%
\input phyzzx
\singlespace
 \frontpagetrue
 \voffset=0.1cm
%%%%%%%%%%%%%%%%%%%%%%%%%%%%% Definitions %%%%%%%%%%%%%%%%%%%%%%%%%%%%%%%%
\def\bg{\hbox{\fourteeni g}}
\def\sbg{\hbox{\twelvei g}}  %% It actually gives the usual metric!
\def\sbeta{\hbox{$\beta \textfont1=\seveni$}}
\def\ssbeta{\hbox{$\beta \textfont1=\fivei$}}
\def\bdelta{\hbox{$\delta \textfont1=\fourteeni$}}
\def\sbdelta{\hbox{$\delta \textfont1=\twelvei$}} %% This is usaul \delta !
\def\cg{\hbox{$\cal G\textfont2=\fourteensy$}}
\def\cq{\hbox{$\cal Q\textfont2=\fourteensy$}}
\def\sqr#1#2{{\vcenter{\vbox{\hrule height.#2pt
        \hbox{\vrule width.#2pt height#1pt \kern#1pt
          \vrule width.#2pt}
        \hrule height.#2pt}}}}
\def\square{\mathchoice\sqr34\sqr34\sqr{2.1}3\sqr{1.5}3}
\def\Square{\mathchoice\sqr67\sqr67\sqr{5.1}3\sqr{1.5}3}
\def\TableOfContentEntry#1#2#3{\relax}
\def\section#1{\par \ifnum\the\lastpenalty=30000\else
   \penalty-200\vskip\sectionskip \spacecheck\sectionminspace\fi
   \global\advance\sectionnumber by 1
   \xdef\sectionlabel{\the\sectionstyle\the\sectionnumber}
   \wlog{\string\section\space \sectionlabel}
   \TableOfContentEntry s\sectionlabel{#1}
   \noindent {\caps\enspace\bf \sectionlabel\quad #1}\par
   \nobreak\vskip\headskip \penalty 30000
   \ifnum\equanumber<0 \else \global\equanumber=0\fi }
\def\eqname#1{\relax \ifnum\equanumber<0
     \xdef#1{{\noexpand\rm(\number-\equanumber)}}%
       \global\advance\equanumber by -1
    \else \global\advance\equanumber by 1
      \xdef#1{{\noexpand\rm(\sectionlabel.\number\equanumber)}} \fi #1}
\def\CQG{Class. Quant. Grav.}
\def\GRG{Gen. Rel. Grav.}
\def\JMP{J. Math. Phys.}
\def\NP{Nucl. Phys.}
\def\PL{Phys. Lett.}
\def\PR{Phys. Rev.}
\def\PRp{Phys. Rep.}
%%%%%%%%%%%%%%%%%%%%%%%%%%%%%%%% References %%%%%%%%%%%%%%%%%%%%%%%%%%%%%%%%%%%
\REF\weyledd{Weyl, H. ``Gravitation und Elektrizit\"at'',
            {\it Preuss. Akad. Wiss. Berlin, Sitz.}\ (1918), 465-480;
            ``Eine neue Erweiterung der Relativit\"atstheorie'',
            {\it Ann. der Phys.}\ {\bf 59}\ (1919), 101-133;
            {\it Raum-Zeit-Materie}, (Springer-Verlag, Berlin, 1$^{\sl st}$
            ed.~1918, 4$^{\sl th}$ ed.~1921). Its English version (of the
            4$^{\sl th}$ ed.) is: {\it Space-Time-Matter}, translated by: H.
            L. Brose (Dover Publications, New York, 1$^{\sl st}$ ed.~1922, reprinted
            1950);
            ``\"Uber die physikalischen Grundlagen der erweiterten
            Relativit\"atstheorie'' \journal Phys. Zeitschr.&22 (21) 473-480;\nextline
            Eddington, A., {\it The Mathematical Theory of Relativity},
           (Cambridge University Press, 1$^{\sl st}$ ed.~1923, 2$^{\sl nd}$
           ed.~1924, Chelsee Publishing Co., New York, 1975).}
\REF\fard{Farhoudi, M. {\it Non--linear Lagrangian Theories of
          Gravitation}, (Ph.D. Thesis, Queen Mary \& Westfield College,
          University of London, 1995).}
\REF\wheb{Wheeler, J. A., {\it Einstein's Vision},
          (Springer-Verlag, Berlin, 1968).}
\REF\bidabos{Birrell, N. D. \& Davies, P. C. W., {\it Quantum
             Fields in
             Curved Space}, (Cambridge University Press, 1982);\nextline
             Buchbinder, I. L., Odintsov, S. D. \& Shapiro, I. L., {\it
             Effective Action in Quantum Gravity}, (Institute of Physics
             Publishing, Bristol and Philadelphia, 1992).}
\REF\utde{Utiyama, R. \& DeWitt, B. S. ``Renormalization of a
          classical
          gravitational interacting with quantized matter fields''
          \journal\JMP&3 (62) 608-618.}
\REF\stela{Stelle, K. S. ``Renormalization of higher derivative
           quantum gravity'' \journal\PR&D16 (77) 953-969.}
\REF\frtsa{Fradkin, E. S. \& Tseytalin, A. A. ``Renormalizable
           asymptotically free quantum theory of gravity'' \journal\NP&B201
           (82) 469-491.}
\REF\stelb{Stelle, K. S. ``Classical gravity with higher
           derivatives'' \journal\GRG&9 (78) 353-371.}
\REF\zwzu{Zwiebach, B. ``Curvature squared terms and string
          theories''
          \journal\PL&156B (85) 315-317;\nextline
          Zumino, B. ``Gravity theories in more than four dimensions''
          \journal\PRp&137 (86) 109-114.}
\REF\lovedc{Lovelock, D. ``The Einstein tensor and its
            generalizations''
            \journal\JMP&12 (71) 498-501; ``The four dimensionality of space
            and the Einstein tensor'' \journal\JMP&13 (72)
            874-876.}
\REF\brig{Briggs, C. C. ``Some possible features of general
          expressions for Lovelock tensors and for the coefficients of Lovelock
          Lagrangians up to the $15^{\sl th}$ order in curvature (and
          beyond)'', {\it gr-qc/9808050}.}
\REF\mtw{Misner, C. W., Thorne, K. S. \& Wheeler, J. A., {\it
         Gravitation},
         (W. H. Freeman \& Company, San Francisco, 1973).}  %%, p 128
\REF\ish{Isham, C. J. ``Quantum gravity'' in
         Proc. 11$^{\sl th}$ {\it General Relativity and Gravitation},
         Stockholm, 1986, Ed. M. A. H. MacCallum (Cambridge University
         Press, London, 1987), pp. 99-129.}
\REF\fare{Farhoudi, M. ``On higher order gravities, their analogy
          to GR, and dimensional dependent version of Duff's trace anomaly
          relation'', {\it \GRG}\ {\bf 38}\ (2006), 1261-1284.}
\REF\noodts{Nojiri, S. \& Odintsov, S. D. ``Introduction to
            modified gravity and gravitational alternative for dark
            energy'', {\it Int. J. Geom. Meth. Mod. Phys.}\ {\bf 4}\ (2007), 115-146.}
\REF\farc{Farhoudi, M. ``Classical trace anomaly'' \journal
          Int. J. Mod. Phys. &D14 (2005) 1233-1250.}
\REF\goss{Gottl\"ober, S., Schmidt, H.--J. \& Starobinsky, A. A.
          ``Sixth--order gravity and conformal transformations''
          \journal\CQG&7 (90) 893-900.}
\REF\berm{Berkin, A. L. \& Maeda, K. ``Effects of $R^3$ and
          $R\,\Square\, R$
          terms on $R^2$ inflation'' \journal\PL&B245 (90) 348-354.}
\REF\buchdcbickstma{Buchdahl, H. A. ``On the gravitational field
                   equations
                   arising from the square of a Gaussian curvature''
                   \journal Nuovo Cim.&23 (62) 141-157;\nextline
                   Bicknell, G. V. ``Non--viability of gravitational theory
                   based on a quadratic Lagrangian'' \journal J. Phys. A:
                   Math. Nucl. Gen.&7 (74) 1061-1069;\nextline
                   Stelle, K. S. ``Classical gravity with higher
                   derivatives''
                   \journal\GRG&9 (78) 353-371;\nextline
                   Maluf, W. ``Conformal invariance and torsion in general
                   relativity'' \journal\GRG&19 (87) 57-71.}
\REF\dadhich{Dadhich, N. ``On the derivation of the gravitational
             dynamics'', {\it gr-qc/0802.3034}.}
\REF\farf{Farhoudi, M. ``New derivation of Weyl invariants in six
          dimensions'', work in progress.}
\REF\mffabsirzbacohiow{Magnano, G., Ferraris, M. \& Francaviglia,
                       M. ``Non--linear gravitational Lagrangians''
                       \journal\GRG&19 (87) 465-479;
                       ``Legendre transformation and dynamical
                       structure of higher derivative gravity''
                       \journal\CQG&7 (90) 557-570;\nextline
                       Sirousse Zia, H. ``Singularity theorems and
                       the [general relativity + additional matter
                       fields] formulation of metric theories of
                       gravitation'' \journal\GRG&26 (94) 587-597;\nextline
                       Barrow, J. D. \& Cotsakis, S. ``Inflation and the conformal
                       structure of higher order gravity theories''
                       \journal\PL&B214 (88) 515-518;\nextline
                       Hindawi, A., Ovrut, B. A. \& Waldram, D. ``Non--trivial
                       vacua in higher derivative gravitation''
                       \journal\PR&D53 (96) 5597-5608.}
\REF\farg{Farhoudi, M. ``{\it Lovelock metric} as a generalized
          metric'', work in progress.}
\REF\dufa{Duff, M. J. ``Observations on conformal anomalies''
          \journal\NP&B125 (77) 334-348.}
\REF\noscenozkmts{Boulware, D. G. \& Deser, S. ``String generated
                  gravity models'', {\it Phys. Rev. Lett.}\ {\bf 55}\ (1985),
                  2656-2660;\nextline
                  Boulware, D. G. \& Deser, S. ``Effective
                  gravity theories with dilatons'', {\it Phys. Lett.}\ {\bf 175B}\
                  (1986), 409-412;\nextline
                  Nojiri, S., Odintsov, S. D. \& Sasaki, M.
                 ``Gauss--Bonnet dark energy'', {\it \PR}\ {\bf
                 D71}\ (2005), 123509;\nextline
                 Cognola, G., Elizalde, E., Nojiri, S., Odintsov, S. D. \& Zerbini,
                 S. ``String--inspired Gauss--Bonnet gravity reconstructed from the
                 universe expansion history and yielding the transition from matter
                 dominance to dark energy'', {\it hep-th/0611198};\nextline
                 Koivisto, T. \& Mota, D. F. ``Cosmology and astrophysical
                 constraints of Gauss--Bonnet dark energy'', {\it \PL}\ {\bf B644}\
                 (2007), 104-108;\nextline
                 Tsujikawa, S. \& Sami, M. ``String--inspired cosmology: a late time
                 transition from scaling matter era to dark energy universe caused
                 by a Gauss--Bonnet coupling'', {\it JCAP}\ {\bf 01}\ (2007), 006.}
\REF\stttnos{Sami, M., Toporensky, A., Tretjakov, P. V. \&
            Tsujikawa, S. ``The fate of (phantom) dark energy universe with string curvature
            corrections'', {\it \PL}\ {\bf B619}\ (2005), 193-200;\nextline
            Nojiri, S., Odintsov, S. D. \& Sami, M. ``Dark energy
            cosmology from higher--order string--inspired gravity and its
            reconstruction'', {\it \PR}\ {\bf D74}\ (2006), 046004.}
\REF\ctsami{Calcagni, G., Tsujikawa, S. \& Sami, M. ``Dark energy
           and cosmological solutions in second--order string gravity'', {\it
           \CQG}\ {\bf 22}\ (2005), 3977-4006.}
%%%%%%%%%%%%%%%%%%%%%%%%%%%%%%%%%%%%%%%%%%%%%%%%%%%%%%%%%%%%%%%%%%%%%%%%%%%%%%%%%
\bigskip
\titlestyle{\bf Lovelock Tensor as Generalized Einstein Tensor}
\bigskip
\centerline{\twelvepoint Mehrdad Farhoudi\foot{This work was
                                               partially
                                               supported by a grant
                                               from the MSRT/Iran.}}
\medskip
\centerline{\tenit Department of Physics, Shahid Beheshti
University, G. C., Evin, Tehran 19839, Iran} \centerline{\tenit
e--mail: m-farhoudi@sbu.ac.ir}
\medskip
\centerline{\ninerm May 21, 2008}
\bigskip
\medskip
\centerline{\bf Abstract}
{\narrower\smallskip \noindent\tenrm\singlespace We show that the
splitting feature of the Einstein tensor, as the first term of the
Lovelock tensor, into two parts, namely the Ricci tensor and the
term proportional to the curvature scalar, {\tenit with} the trace
relation between them is a common feature of {\tenit any} other
homogeneous terms in the Lovelock tensor. Motivated by the
principle of general invariance, we find that this property can be
generalized, with the aid of a {\tenit generalized trace operator}
which we define, for any {\tenit inhomogeneous} Euler--Lagrange
expression that can be spanned linearly in terms of homogeneous
tensors. Then, through an application of this generalized trace
operator, we demonstrate that the Lovelock tensor analogizes the
mathematical form of the Einstein tensor, hence, it represents a
generalized Einstein tensor. Finally, we apply this technique to
the scalar Gauss--Bonnet gravity as an another version of
string--inspired gravity.
\smallskip}
%\bigskip
\medskip
\noindent
 {\ninerm PACS number:} {\tenpoint $04.20.-q$ ; $04.50.+h$} \nextline
 {\nineit Keywords}: {\ninerm Higher Order Gravity; Lovelock \& Einstein Tensors;
           Trace Operator.}
\section{\bf Introduction}
\indent
 Among scalar Lagrangians, field equations based on a
Lagrangian quadratic in the curvature tensor have had a long
history in the theory of gravitation~[\weyledd]. Perhaps a
legitimate mathematical motivation to examine gravitational
theories built on non--linear Lagrangians has been the
phenomenological character of the Einstein theory, that is the
dependence of the Einstein tensor and Lagrangian on the
derivatives of the metric, which leaves room for such amendments
and the dimension~[\fard]. Actually, the Einstein Lagrangian is
not the most general second order Lagrangian allowed by the
principle of general invariance, and indeed, through this
principle the latter generalization can be performed up to {\it
any} order, and a general scalar Lagrangian is a higher derivative
Lagrangian.
\par
Despite Einstein's gravitational theory successes, its main
difficulties become manifest when the curvature of space--time is
not negligible, e.g., in the very early universe with distances of
the order of Planck's length, where an Euclidean topological
structure is quite unlikely. At such distances, even the
fluctuations of quantum gravitation will be extremely violent and
probably produce an ever changing, dynamic topology~[\wheb]. This
perhaps allows non--linear gravitational Lagrangians to be
considered as alternative theories.
\par
Nowadays, it is also well known that Einstein's gravity when
treated as a fundamental quantum gravity
leads\footnote{\diamond}{See, for example, Refs.~[\bidabos] and
references therein.}\
 to a non--renormalizable theory. In order to permit renormalization
of the divergences, quantum gravity has indicated that the
Einstein--Hilbert action should be enlarged by the inclusion of
higher order curvature terms~[\utde]. In fact, it has been
shown~[\bidabos] that the Lagrangian $L={1
\over\kappa^2}\bigl(R+\alpha R^2+\beta
    R_{\mu\nu}R^{\mu\nu}\bigr)$\rlap,\footnote{\star}{$\alpha$, $\beta$ and
 $\kappa^2\equiv {16\pi G\over c^4}$ are constants, and
 field equations are shown as
 $G^{(\sl gravitation)}_{\alpha\ssbeta}={1\over 2}\kappa^2\,
 T_{\alpha\ssbeta}$, where the definition $\sbdelta\left(L_{\rm m}\sqrt{-g}\,\right)
 \equiv -{1\over 2}\sqrt{-g}\ T_{\alpha\sbeta}\, \sbdelta
 \sbg^{\alpha\sbeta}$ is used.}\
which, by the Gauss--Bonnet~({\bf GB}) theorem, is the most
general quadratic Lagrangian in, and up to, four dimensions and
has the required Newtonian limit, solves the renormalization
problem and is multiplicatively renormalizable~[\stela] and
asymptotically free~[\frtsa]; however it is~not
unitary\footnote{\ddag}{A characteristic property of unitarity is
the scalar product, or norm, invariance.}\
 within usual perturbation theory~[\stelb].
Actually, its particle spectrum contains a further massive scaler
spin--two ghost, which has either negative energy or a negative
norm, and the existence of negative energy excitations in a model
leads to causality violation~[\stelb]. However, the lack of
unitarity is the main reason for not considering higher order
gravities as the best candidate for the quantum gravity
description.
\par
The theory of superstrings, in its low energy limits, also
suggests the above inclusions, and in order to be ghost--free it
is shown~[\zwzu] that it must be in the form of dimensionally
continued GB densities, that is the Lovelock
Lagrangian~[\lovedc,\brig],
\vadjust{\nobreak\foot{Our conventions are a metric of signature
$+2$,\
$R^{\mu}{}_{\nu\alpha\ssbeta}=-\Gamma^{\mu}{}_{\nu\alpha,\,\ssbeta}+\cdots$,
and $R_{\mu\nu}\equiv R^{\alpha}{}_{\mu\alpha\nu}$.}}
$$
{\cal L}={1 \over\kappa^2}\sum_{0<n<{D\over 2}}\,{1\over 2^n}\,
   c_n\,\bdelta^{\alpha_1\ldots\alpha_{2n}}_{\sbeta_1\ldots\sbeta_{2n}}\,
   R_{\alpha_1\alpha_2}{}^{\sbeta_1\sbeta_2}\cdots R_{\alpha_{2n-1}\,
   \alpha_{2n}}{}^{\sbeta_{2n-1}\,\sbeta_{2n}}
   \equiv\sum_{0<n<{D\over 2}}\, c_n\, L^{(n)}\ \rlap,\footattach\eqno\eq
$$
where we set $c_1\equiv 1$ and the other $c_n$ constants to be of
the order of Planck's length to the power $2(n-1)$, for the
dimension of $\cal L$ to be the same as $L^{(1)}$. Symbol
$\bdelta^{\alpha_1\ldots\alpha_p}_{\sbeta_1\ldots\sbeta_p}$ is the
generalized Kronecker delta symbol, see, e.g., Ref.~[\mtw], which
is identically zero if $p>D$. The maximum value of $n$ is related
to space--time dimension, $D$, by
$$
n_{_{\rm
max.}}\!=\cases{{D\over 2}-1&{\rm even} $D$\cr
         {{D-1}\over 2}&{\rm odd} $D$\ .\cr}\eqno\eq
$$
An important aspect of this suggestion is~[\ish] that it does not
arise in attempts to quantize gravity. The above ghost--free
property, and the fact that the Lovelock Lagrangian is the most
general second order Lagrangian which, the same as the
Einstein--Hilbert Lagrangian, yields the field equations as {\it
second} order equations, have stimulated interests in Lovelock
gravity and its applications in the literature, see, e.g.,
Refs.~[\fare,\noodts] and references therein. The Lovelock
Lagrangian obviously reduces to the Einstein--Hilbert Lagrangian
in four dimensions and its second term is the GB invariant,
$L^{(2)}={1 \over\kappa^2}\bigl(R^2-4
R_{\mu\nu}R^{\mu\nu}+R_{\alpha\sbeta\mu\nu}\,
 R^{\alpha\sbeta\mu\nu}\bigr)$.
\par
We have noticed that each term of the Lovelock tensor, that is
$G^{(n)}_{\alpha\sbeta}$, where the Lovelock tensor
is~[\lovedc,\brig]
\vadjust{\nobreak\foot{We have neglected the cosmological term,
and $G^{(1)}_{\alpha\ssbeta}=G_{\alpha\ssbeta}$ that is, the
Einstein tensor.}}
$$
\eqalign{\cg_{\alpha\sbeta}
   &=-\sum_{0<n<{D\over 2}}\, {1\over 2^{n+1}}\, c_n\, \bg_{\alpha\mu}\>
    \bdelta^{\mu\alpha_1\ldots\alpha_{2n}}_{\sbeta\sbeta_1\ldots\sbeta_{2n}}
    \, R_{\alpha_1\alpha_2}{}^{\sbeta_1\sbeta_2}\cdots
    R_{\alpha_{2n-1}\,\alpha_{2n}}{}^{\sbeta_{2n-1}\,\sbeta_{2n}}\crr
   &\equiv\sum_{0<n<{D\over 2}}\, c_n\, G^{(n)}_{\alpha\sbeta}
    \ \rlap,\footattach\cr}\eqno\eq
$$
has also the following remarkable properties. We mean, each term
of the $G^{(n)}_{\alpha\sbeta}$ can be rewritten in a form that
analogizes the Einstein tensor with respect to the Ricci and the
curvature scalar tensors, namely
$G^{(n)}_{\alpha\sbeta}=R^{(n)}_{\alpha\sbeta}-{1\over
2}g_{\alpha\sbeta}\, R^{(n)}$\ where
$$
R^{(n)}_{\alpha\sbeta}\equiv {n\over 2^n}\,\bdelta^{\alpha_1\alpha_2%
   \ldots\alpha_{2n}}_{\alpha\ \sbeta_2\ldots\sbeta_{2n}}\,
   R_{\alpha_1\alpha_2\sbeta}{}^{\sbeta_2}\, R_{\alpha_3\alpha_4}
   {}^{\sbeta_3\sbeta_4}\cdots R_{\alpha_{2n-1}\,\alpha_{2n}}
   {}^{\sbeta_{2n-1}\,\sbeta_{2n}}\ ,\eqno\eq
$$
$$
R^{(n)}\equiv {1\over 2^n}\,\bdelta^{\alpha_1\ldots\alpha_{2n}}_{\sbeta_1
    \ldots\sbeta_{2n}}\, R_{\alpha_1\alpha_2}{}^{\sbeta_1\sbeta_2}\cdots
    R_{\alpha_{2n-1}\,\alpha_{2n}}{}^{\sbeta_{2n-1}\,\sbeta_{2n}}
    \ ,\eqno\eq
$$
$R^{(1)}_{\alpha\sbeta}\equiv R_{\alpha\sbeta}$ and $R^{(1)}\equiv
R$.
\par
The proof of these can easily be done using the definition of the
generalized Kronecker delta symbol and the properties of the
Riemann--Christoffel tensor. An alternative and more basic
approach is to notice that it is what one gets in the process of
varying the action, $\bdelta\int L^{(n)}\sqrt{-g}\, d^D\!x$, where
its Euler--Lagrange expression will be
$$
{\bdelta L^{(n)}\over \bdelta\bg^{\alpha\sbeta}}-{1\over 2}
   \bg_{\alpha\sbeta}\, L^{(n)}\equiv {1\over\kappa^2}\,
   G^{(n)}_{\alpha\sbeta}\ .\eqno\eq
$$
Hence, from relations~(1.1), (1.4) and (1.5), one can easily show
that
$$
{\bdelta L^{(n)}\over
\bdelta\bg^{\alpha\sbeta}}={1\over\kappa^2}\,
R^{(n)}_{\alpha\sbeta} \qquad{\rm and}\qquad
L^{(n)}={1\over\kappa^2}\, R^{(n)}\ .\eqno\eq
$$
Almost the same procedure has, perhaps, been carried by
Lovelock~[\lovedc], but he then proceeded from this to derive
equation~(1.3).
\par
Before we continue, we should indicate that one does not need
necessarily to restrict oneself to relations~(1.4), (1.5) and
(1.7) in any kind of generic field equations and may take any term
in the final result of ${\delta L{}\ {}\over \delta
g^{\alpha\ssbeta}}$ which appears to be a scalar multiple of
$\bg_{\alpha\sbeta}$ out of it, see Ref.~[\farc].
\par
Although the above derivation is straightforward, what is not so
obvious at the first sight is that there also exists a relation
between $R^{(n)}_{\alpha\sbeta}$ and $R^{(n)}$ analogous to that
which exists between the Ricci tensor and the curvature scalar,
namely
$$
{1\over n}\,{\rm trace}\, R^{(n)}_{\alpha\sbeta}=R^{(n)}\
,\eqno\eq
$$
where the trace means the standard contraction of any
two indices that is, for example, ${\sl trace}\, A_{\mu\nu}\equiv
\bg^{\alpha\sbeta}A_{\alpha\sbeta}$.
\par
Hence, the splitting feature of the Einstein tensor, as the first
term of the Lovelock tensor, into two parts {\it with} the trace
relation between them is a common feature of {\it any} other terms
in the Lovelock tensor, in which each term alone is a homogeneous
Lagrangian. Indeed, this property has been resulted through the
variation procedure. Thus, motivated by the principle of general
invariance, one also needs to consider what might happen if the
Lagrangian under consideration, and hence its relevant
Euler--Lagrange expression is an {\it inhomogeneous} tensor, as
for example, the (whole) Lovelock Lagrangian, ${\cal L}$, which is
constructed of terms with a mixture of different orders.
\par
In this case, the relevant Euler--Lagrange expression can easily
be written by analogy with the form of $G^{(n)}_{\alpha\sbeta}$,
for example, $\cg_{\alpha\sbeta}=\Re_{\alpha\sbeta}-{1\over
2}g_{\alpha\sbeta}\,\Re$\ where
$$
\Re_{\alpha\sbeta}\equiv
    \sum_{0<n<{D\over 2}}\, c_n\, R^{(n)}_{\alpha\sbeta}
    \qquad{\rm and}\qquad
    \Re\equiv \sum_{0<n<{D\over 2}}\, c_n\, R^{(n)}\ .\eqno\eq
$$
But a similar relation to equation~(1.8) does not apparently hold
between $\Re_{\alpha\sbeta}$ and $\Re$ due to the factor ${1\over
n}$.
\par
To overcome this issue, in the next section we will introduce a
generalized trace as an extra mathematical tool for Riemannian
manifolds, which will slightly alter the original form of the
trace relation and modify it sufficiently to enable us to deal
with the above difficulty. Then, in Section~3, we will consider
the case of the inhomogeneous Lovelock tensor, where also some
discussions will be presented, and in the last section, we will
apply our technique to a more interested case of the scalar GB
gravity.
\section{\bf Generalized Trace}
\indent
In this section, we will define a {\it generalized} trace, which we will
denote by {\it Trace}, for tensors whose components are homogeneous
functions of the metric and its derivatives.
\par
But first, as either of $\bg_{\mu\nu}$ or $\bg^{\mu\nu}$ can be
chosen as a base for counting the homogeneity degree numbers, we
will choose, without loss of generality and as a convention, the
homogeneity degree number ({\bf HDN}) of $\bg^{\mu\nu}$ as [{\bf
+}1]; hence, the HDN of $\bg_{\mu\nu}$ will be $[-1]$ since
$\bg^{\mu\nu}\,\bg_{\mu\alpha}=\bdelta^\nu_\alpha$. So, as
contravariant and covariant tensors are mapped into each other in
a one--to--one manner by the metric, their HDNs are different by
$\pm 2$. Similarly, we will choose the HDN of
$\bg^{\mu\nu}{}_{,\,\alpha}$ as [{\bf +}1]. Therefore, from
$\bg_{\alpha\sbeta ,\,\rho}=-\bg_{\alpha\mu}\,\bg_{\sbeta\nu}\,
                                          \bg^{\mu\nu}{}_{,\,\rho}$,
the HDN of $\bg_{\mu\nu,\,\alpha}$ will be $[-1]$ as well. To
specify the HDNs of higher derivatives of the metric, one may
consider $\partial_\alpha$ as if with the HDN of zero, and hence
for $\partial^\alpha
\bigl(=\bg^{\alpha\sbeta}\,\partial_{\sbeta}\bigr)$ as if with
[+1]. By the very elementary property of homogeneous functions,
the HDN of a term consists of cross functions is obviously found
by adding the HDN of each of the cross functions. For convenience,
the HDNs, $h$, of some homogeneous functions of the metric and its
derivatives are given in Table~1, and whenever necessary, we will
show the HDN of a function in brackets attached to the upper left
hand side, e.g., ${}^{[+1]}\bg^{\mu\nu}$ and
${}^{[-1]}\bg_{\mu\nu}$.
\par
Now, for a general $\Bigl({N\atop M}\Bigr)$ tensor, e.g.,
$A^{\alpha_1\ldots\alpha_N}{}_{\sbeta_1\ldots\sbeta_M}$, which is
a homogeneous function of degree $h$ with respect to the metric
and its derivatives, we define\foot{These definitions match our
HDN conventions of ${}^{[+1]}\sbg^{\mu\nu}$ and
${}^{[+1]}\sbg^{\mu\nu}{}_{,\alpha}$ to comply with our needs.}\
$$
{\rm Trace}\, {}^{[h]}A^{\alpha_1\ldots\alpha_N}\!{}_{\sbeta_1\ldots\sbeta_M}
    \!:=\!\cases{\!{1\over h-{N\over 2}+{M\over 2}}\, {\rm trace}\,
     {}^{[h]}A^{\alpha_1\ldots\alpha_N}\!{}_{\sbeta_1\ldots\sbeta_M}&\quad
     $\!${\rm when} $h\!-\!{N\over 2}\!+\!{M\over 2}\!\not=\!0$\crr
     \!{\rm trace}\, {}^{[h]}A^{\alpha_1\ldots\alpha_N}{}
     _{\sbeta_1\ldots\sbeta_M}&\quad $\!${\rm when}
     $h\!-\!{N\over 2}\!+\!{M\over 2}\!=\!0$\ .\cr}\eqno\eq
$$

\nextline \centerline{\tenpoint
          {\bf Table 1}:\ The HDNs of useful homogeneous functions.}
{\tenpoint
$$
\vbox{\offinterlineskip
  \hrule
  \halign{&\vrule#&\strut\quad\hfil#\hfil\quad
          &\vrule#&\strut\quad\hfil#\hfil\quad&\vrule#\cr
    height2pt&\omit&&\omit&\cr
    &{\bf Function}&&{\bf The HDN}&\cr
    height2pt&\omit&&\omit&\cr
    \noalign{\hrule}
    height2pt&\omit&&\omit&\cr
    &$\sbg^{\mu\nu}$ ({\rm our convention})&&+1&\cr
    height2pt&\omit&&\omit&\cr
    &$\sbg_{\mu\nu}$&&$-1$&\cr
    height2pt&\omit&&\omit&\cr
    &$\sbg^{\mu\nu}{}_{,\,\alpha}$ ({\rm our convention})&&+1&\cr
    height2pt&\omit&&\omit&\cr
    &$\sbg^{\mu\nu,\,\alpha}$&&+2&\cr
    height2pt&\omit&&\omit&\cr
    &{\rm Operator}\ $\partial_\alpha$&&{\rm as\ if}\ \ 0&\cr
    height2pt&\omit&&\omit&\cr
    &{\rm Operator}\ $\partial^\alpha$&&{\rm as\ if}\ \ +1&\cr
    height2pt&\omit&&\omit&\cr
    &$\sbg_{\mu\nu,\,\alpha}$&&$-1$&\cr
    height2pt&\omit&&\omit&\cr
    &$\sbg^{\mu\nu}{}_{,\,\alpha\ssbeta\ldots}$&&+1&\cr
    height2pt&\omit&&\omit&\cr
    &$\sbg_{\mu\nu,\,\alpha\ssbeta\ldots}$&&$-1$&\cr
    height2pt&\omit&&\omit&\cr
    &$g\equiv \det (\sbg_{\mu\nu})$&&$-D$&\cr
    height2pt&\omit&&\omit&\cr
    &$\Gamma^\alpha{}_{\mu\nu}$&&0&\cr
    height2pt&\omit&&\omit&\cr
    &$R^\alpha{}_{\ssbeta\mu\nu}$&&0&\cr
    height2pt&\omit&&\omit&\cr
    &$R_{\alpha\ssbeta\mu\nu}$ \&\ $C_{\alpha\ssbeta\mu\nu}$&&$-1$&\cr
    height2pt&\omit&&\omit&\cr
    &$R_{\mu\nu}$&&0&\cr
    height2pt&\omit&&\omit&\cr
    &$R^{\mu\nu}$&&+2&\cr
    height2pt&\omit&&\omit&\cr
    &$R$&&+1&\cr
    height2pt&\omit&&\omit&\cr
    &$L^{(n)}$ \&\ $R^{(n)}$&&$+n$&\cr
    height2pt&\omit&&\omit&\cr
    &$R^{(n)}_{\mu\nu}$ \&\ $G^{(n)}_{\mu\nu}$&&$n-1$&\cr
    height2pt&\omit&&\omit&\cr}
  \hrule}
$$ }
\indent
 Contravariant and covariant components of a tensor
obviously have different HDNs, however, by the above definition,
the equality of their {\sl Traces} is still retained. For example,
${\sl Trace}\, A_{\mu\nu}={\sl Trace}\, A^{\mu\nu}\equiv A$\
regardless of what the HDN, just as for the {\sl trace} operator
that is, ${\sl trace}\, A_{\mu\nu}={\sl trace}\, A^{\mu\nu}\equiv
A^\alpha{}_\alpha$. Note that using our definition it follows that
$$
\cases{A={1\over h+1}\, A^{\alpha}{}_{\alpha}&\quad {\rm for}
$h\not=-1$\cr
       A=A^{\alpha}{}_{\alpha}&\quad {\rm for} $h=-1$\ ,\cr}\eqno\eq
 $$
where the factor of ${1\over h+1}$ has entered because we have
taken ${}^{[h]}A_{\mu\nu}$, and therefore used the fact that the
HDNs of both $A$ and $A^{\alpha}{}_{\alpha}$ are $[h+1]$.
\par
In general, the generalized trace has, by its definition, all of
the properties of the usual trace, especially its invariance under
a similarity transformation (for similar tensors) if the
transformation does not change the HDN of the tensor, and its
basis independence for linear operators on a finite dimensional
Hilbert space. However, as we will show in the following, it
cannot act as a linear operator\foot{That is, for example, \ ${\sl
trace}\,\Bigl(a_1 A_{\mu\nu}+a_2 B_{\mu\nu}\Bigr)=a_1\,{\sl
trace}\, A_{\mu\nu}+a_2\, {\sl trace}\, B_{\mu\nu}$.}\ when the
coefficients of linearity themselves are any scalar homogeneous
functions of degree $h'\not=0$.
\par
By the definition of the {\it Trace}, for cases when $h'\not=0$,
we have, for example,
$$
\eqalign{{\rm Trace}\,\Bigl({}^{[h']}C\, {}^{[h]}A_{\mu\nu}\Bigr)
          &={1\over h'+h+1}\,{\rm trace}\,\Bigl({}^{[h']}C\,
           {}^{[h]}A_{\mu\nu}\Bigr)\qquad {\rm for}\ h'+h\not=-1\cr
          &={{}^{[h']}C\over h'+h+1}\,{\rm trace}\,
           {}^{[h]}A_{\mu\nu}\ ,\cr}
 $$
and using the definition once again, we get
$$
{\rm Trace}\,\Bigl({}^{[h']}C\, {}^{[h]}A_{\mu\nu}\Bigr)=
    \cases{{h+1\over h'+h+1}\, {}^{[h']}C\,{\rm Trace}\,
           {}^{[h]}A_{\mu\nu}&\quad {\rm for} $h\not=-1$\crr
           {1\over h'}\, {}^{[h']}C\,{\rm Trace}\,
           {}^{[h]}A_{\mu\nu}&\quad {\rm for} $h=-1$\ .\crr}\eqno\eq$$
Alternatively, we obtain
$$
\eqalign{{\rm Trace}\,\Bigl({}^{[h']}C\, {}^{[h]}A_{\mu\nu}\Bigr)
        &={}^{[h']}C\,{\rm trace}\, {}^{[h]}A_{\mu\nu}\ \qquad\quad
        {}\quad\
         {\rm for}\ h'+h=-1\cr
        &=\bigl(h+1\bigr)\, {}^{[h']}C\,{\rm Trace}\,
         {}^{[h]}A_{\mu\nu}\quad {\rm for}\ h\not=-1\ .\cr}\eqno\eq
 $$
Obviously, these extra factors can be made equal to one, only when
$h=-1$ and $h'=+1$, or when $h=0$ and $h'=-1$, as in second
equation of~(2.3) or in equation~(2.4), respectively.
\par
Therefore, due to these extra factors the {\it Trace} is not a
linear operator as mentioned above. \ $\square$
\par
It is necessary to emphasize that for our immediate purposes,
which led us initially to define a generalized trace, the {\it
Trace} indeed has the distributive law of the usual trace in the
cases when there are either no coefficients of linearity, or when
coefficients are included with their associated tensors, and or
when coefficients are meant to be scalars with $h'=0$.
\par
To justify the way that we have defined a generalized trace,
other than that it satisfies our need for dealing with inhomogeneous
Lagrangians, one can show that this definition also has a link with Euler's
theorem for homogeneous functions.
\par
Suppose $A(g^{\mu\nu})$ is a homogeneous scalar function of degree
$[h]$, i.e. $A(\lambda g^{\mu\nu})=\lambda^h\, A(g^{\mu\nu})$.
Euler's theorem states that $\bg^{\mu\nu}\, {\partial\, A{}\;\
{}\over \partial\sbg^{\mu\nu}}=h\, A$. As a rough and ready
argument, define ${\partial\, A{}\;\ {}\over
\partial\sbg^{\mu\nu}}\equiv A_{\mu\nu}$, where $A_{\mu\nu}$ is of
degree $h-1$, then $\bg^{\mu\nu}\, A_{\mu\nu}$ denotes its usual
trace. Also, define $A\equiv {\sl Trace}\, A_{\mu\nu}$. Then, from
Euler's theorem, one can derive ${\sl
trace}\,{}^{[h-1]}A_{\mu\nu}=h\,{\sl Trace}\, A_{\mu\nu}$, or
$$
{\rm Trace}\, {}^{[h-1]}A_{\mu\nu}={1\over h}\, {\rm trace}\,
       A_{\mu\nu}\qquad {\rm when}\ h\not=0\ ,\eqno\eq
 $$
which is the same as our definition~(2.1). When $h=0$, which means
that $A$ does not depend on the metric and its derivatives,
Euler's theorem is trivial, that is ${\partial\, A{}\;\ {}\over
\partial\sbg^{\mu\nu}}=0$. Therefore, the best and consistent
choice is to make no distinction between {\it Trace} and the trace
for ${}^{[0]}A$.
\par
On the other hand, using Table~1, it is straightforward to
relate~[\fare] the orders $n$ in any Lagrangian, as in $L^{(n)}$,
that represents its HDN\rlap.\foot{This choice is as to be
consistent with our HDN conventions of ${}^{[+1]}\sbg^{\mu\nu}$
and ${}^{[+1]}\sbg^{\mu\nu}{}_{,\,\alpha}$, and with our
definition of {\it Trace}.}\ So, one may refer to Lagrangians with
their HDNs rather than their orders. As an immediate efficiency,
consider the following example. In order to amend the Lagrangian
of {\it sixth order gravity}~[\goss], Berkin {\it et al}~[\berm]
discussed that the Lagrangian term of $R\,\Square\, R$ is a third
order Lagrangian based on the dimensionality scale, for two
derivatives are dimensionally {\it equivalent} to one
Riemann--Christoffel tensor or any one of its contractions.
However, it can be better justified on account of the above
regard, since it has the HDN {\it three}.
\par
Using the above aspect, the special case of $h=0$, similar to
relation~(1.1), corresponds to $c_0\, L^{(0)}\equiv
2\Lambda/\kappa^2$, a constant, which produces the cosmological
term, $-\Lambda\, {}^{[-1]}\bg_{\mu\nu}$ or equivalently
$\Lambda\, {}^{[+1]}\bg^{\mu\nu}\,$, in the field equations.
Hence, the exception value in our definition of generalized trace,
equation~(2.1), for an scalar maybe related to the cosmological
term difficulty, see Ref.~[\fare] for more details. Nevertheless,
with our choice of definition for the generalized trace, one has
$$
{\rm Trace}\, {}^{[+1]}\bg^{\mu\nu}={\rm trace}\, \bg^{\mu\nu}=D
\qquad{\rm and}\qquad {\rm Trace}\, {}^{[-1]}\bg_{\mu\nu}={\rm
trace}\, \bg_{\mu\nu}=D\ .
$$
\par
Finally, as an example, if one applies the definition of
generalized trace on equation~(1.4) a relation similar to
equation~(1.8) will be obtained, but in even more analogous form
for each order, namely
$$
\eqalign{
   &{\rm Trace}\, R^{(1)}_{\alpha\sbeta}
    =R^{(1)}=\kappa^{2}\, L^{(1)}\cr
   &{\rm Trace}\, R^{(2)}_{\alpha\sbeta}
    =R^{(2)}=\kappa^{2}\, L^{(2)}\cr
   &{\rm Trace}\, R^{(3)}_{\alpha\sbeta}
    =R^{(3)}=\kappa^{2}\, L^{(3)}\cr
   &\ \vdots\cr
   &{\rm Trace}\, R^{(n)}_{\alpha\sbeta}
    =R^{(n)}=\kappa^{2}\, L^{(n)}\ ,\cr}\eqno\eq
 $$
where, by equation~(2.2), we generally have
$$
R^{(n)}={1\over n}\, R^{(n)\,\rho}{}_\rho \ ,\eqno\eq
 $$
and, for example, $R^{(1)}=R^{(1)\,\rho}{}_\rho\equiv R$, as
expected.
\section{\bf Inhomogeneous Lovelock Tensor and Discussions}
\indent
 As a simple evident of efficiency of the Trace operator,
we will apply it to the Lovelock tensor.
\par
In the case of inhomogeneous Lovelock Lagrangian, we have shown
that the Lovelock tensor can be written as
$\cg_{\alpha\sbeta}=\Re_{\alpha\sbeta}-{1\over
2}\bg_{\alpha\sbeta}\,\Re$. Furthermore, by substituting for
$R^{(n)}$ from equation~(1.5) and using equation~(1.1), we have
${\cal L}={1\over \kappa^2}\,\Re$ and also by substituting for it
from equation~(2.6), then using the distributivity of the {\it
Trace}, we get $\Re={\sl Trace}\, \Re_{\alpha\sbeta}$, which is
exactly in the same form as the equations of~(2.6).
\par
Hence, in higher dimensional space--times, the Lovelock tensor,
which reduces to the Einstein tensor in four dimensions and its
other useful and interest properties have been summarized in the
introduction, analogizes the mathematical form of the Einstein
tensor as well. Implicitly, in higher order gravities, where the
geometry is represented by the Lovelock tensor, the field
equations can be written as
$$
\cg_{\alpha\sbeta}={1 \over 2}\kappa^2\, T_{\alpha\sbeta}\
.\eqno\eq
 $$
We therefore classify the Lovelock tensor, as a {\it generalized}
Einstein tensor, and call ${\cal L}$, $\Re_{\alpha\sbeta}$ and
$\Re$ the {\it generalized}\ Einstein's gravitational Lagrangian,
{\it generalized}\ Ricci tensor and {\it generalized}\ curvature
scalar, respectively.
\par
Recently, it has been claimed~[\dadhich] that a distinct, but
equivalent, derivation of the gravitational dynamics can be
obtained for a Lovelock--type action from the trace of a
Bianchi--type identity satisfied by a fourth rank tensor which is
a polynomial in curvature. Besides, the trace of such a tensor is
equal to the corresponding Lagrangian. This has been
demonstrated~[\dadhich] separately for the Einstein gravity, and
also for the GB term alone. However, we know that the unity of
physics during its development must be maintained by the {\it
correspondence principle}. That is, in every new physical theory
the previous one is contained as a limiting case. Indeed,
gravitational theories based on a Lagrangian which is {\it only}
purely quadratic in the curvature tensor have been strongly
criticized~[\buchdcbickstma] as nonviable theories. The two main
objections against these Lagrangians are as the metric based on
them does not satisfy the flat space limit at asymptotically large
distances; and disagreement with observations follows when the
matter is incorporated. Therefore, one should demand that
Einstein's gravity must be maintained as a limiting case of
non--linear gravitational Lagrangians. On this ground, if one
wants to perform the approach of Ref.~[\dadhich] for the Einstein
plus the GB term gravity, or in general for the (whole) Lovelock
polynomial terms together, one must employ the Trace operator
instead of the usual trace one, for the same reasons explained in
the introduction.
\par
More applications and evident of usefulness of this Trace operator
have recently been shown in Refs.~[\fare,\farc,\farf]. This is
perhaps the main reason why we have been encouraged to present the
details of its definition, for a wider audience by its
publication.
\par
In this work, our main effort has been to define a generalized
trace as an extra mathematical tool, by which, and as its
application, we have shown that the analogy of the Einstein tensor
can be generalized to any inhomogeneous Euler--Lagrange expression
if it can be spanned linearly in terms of homogeneous tensors,
e.g., the Lovelock tensor. The significant use of this new
operator is that one can apply it to achieve the Lovelock gravity
as a generalization of the Einstein gravity. However, we should
emphasis that by making such a new definition for generalized
trace, we do not really mean to change the essential or inherent
properties of the original Lovelock tensor. The underlying hope
motivating our work is to grasp better insight and understanding
of the properties and abilities of the Lovelock gravity.
\par
Besides the very well known classical successes of Einstein's
theory, the above analogy may be used as a part of a programme to
impose a total analogy of Einstein's gravity on Lovelock's
gravity~[\fard], wherein the latter can then be considered as a
generalized Einstein's gravity. A tentative suggestion that may
relate higher order gravities under some kind of transformation,
e.g., conformal and or Legendre--like transformations, see, for
example, Refs.~[\mffabsirzbacohiow]. Hence, a major task will be
to construct, and hence to achieve, a generalized counterpart for
each essential term used in Einstein's gravity, especially the
metric, for which the task is under investigation~[\farg]. This,
we believe, would give a better view on higher order gravities,
and would also let straightaway to apply the results of Einstein's
theory to Lovelock's theory. At least, such a generalization is of
potential importance as it gives an alternative framework in order
to derive consequences analogous to those obtained in general
relativity for the generalized theories of gravities.
\par
Almost in the same spirit, a further analogy in the mathematical
form of the alternative {\it form} of Einstein's field equations
and the relevant alternative form of Lovelock's field equations
have shown that the price for this analogy is to accept the
existence of the trace anomaly of the energy--momentum tensor even
in classical treatments~[\farc]. Investigation has
indicated~[\fard,\farc] that there is an interesting similarity
between the trace anomaly relation suggested by Duff~[\dufa] and
the constraint relation that coefficients of any generic second
order Lagrangian must satisfy in order to hold the desired
analogy. That is, one may speculate that the origin of Duff's
suggested relation is classically due to the covariance of the
form of Einstein's equations~[\farc]. And, in Ref.~[\fare], a
dimensional dependent version of Duff's trace anomaly relation has
been derived based on this analogy, in where its important
achievements have also been itemized.
\par
Implicitly, one must note that, wherever a term such as
$\bg_{\alpha\sbeta}\, R$ is involved in an equation, its analogous
counterpart, $\bg_{\alpha\sbeta}\, \Re$, may not lead to the
correct corresponding equation in its generalized form. For
example, in a $D$--dimensional space--time, a traceless Ricci
tensor is usually defined as
$$
Q_{\alpha\sbeta}\equiv R_{\alpha\sbeta}-{1\over D}\,\bg_{\alpha\sbeta}
                       \, R\ .\eqno\eq
 $$
Whereas, the corresponding generalized {\it Traceless} Ricci
tensor can only be defined as
$$
\cq_{\alpha\sbeta}\equiv \Re_{\alpha\sbeta}-{1\over D}\,
         \bg_{\alpha\sbeta}\, \sum_{0<n<{D\over 2}}\, n\, c_n\,
         R^{(n)}\ ,\eqno\eq
 $$
and obviously, it is not the covariant form of the former
relation. However, our main concern is the analogy in the
fundamental equations of any theory of gravity.
\par
In the next section we will apply the Trace operator to the scalar
Gauss--Bonnet ({\bf SGB}) gravity.
\section{\bf Scalar Gauss--Bonnet Gravity}
\indent

The Lovelock Lagrangian in its total form has hardly been used in
physical backgrounds. As, we have applied~[\fare,\farc] the Trace
operator for the first plus second terms of the Lovelock
Lagrangian, that is the GB gravity. Also, we have employed this
technique for third order term of the Lovelock Lagrangian~[\farf].
But, despite the successes of the Lovelock gravity, especially the
GB gravity, one may performs to consider also account of scalars,
e.g., in simplest case, dilaton. Indeed, the SGB gravity as an
another version of string--inspired gravity, has been suggested,
Refs.~[\noscenozkmts] and references therein, to be used recently
as a possibility for gravitational dark energy, in order to
explain the observed acceleration of the universe. Also, inclusion
of higher order terms has been considered in Refs.~[\stttnos].
This scenario exhibits several features of cosmological interest
for late universe~[\noodts,\noscenozkmts,\stttnos] and
applications in the early universe, see, e.g., Ref.~[\ctsami] and
references therein. Hence, it may be of more interest and or
useful for physicists to apply the Trace operator to the SGB
gravity, where this section is devoted to.
\par
Typically the low--energy limit of the string theory characterizes
scalar fields and their couplings to various curvature terms. We
consider the effective action given by
$$
S=\int \left[L^{\rm (SGB)}+L_{\rm m}\right]\sqrt{-g}\, d^D\!x\
,\eqno\eq
 $$
where $L^{\rm (SGB)}\equiv L^{(1)}+L^{(\phi)}+L^{(0)}+f(\phi)
L^{(2)}$, and where $L^{(1)}={1\over\kappa^2}R$ the usual
Einstein--Hilbert Lagrangian, $L^{(\phi)}\equiv
-{1\over\kappa^2}{\gamma\over
2}\partial_{\mu}\phi\,\partial^{\mu}\phi$ with $\gamma=\pm 1$ for
the canonical scalar and or phantom, $L^{(0)}\equiv
-{1\over\kappa^2}V(\phi)$ and $L^{(2)}={1
\over\kappa^2}\bigl(R^2-4
R_{\mu\nu}R^{\mu\nu}+R_{\alpha\sbeta\mu\nu}\,
 R^{\alpha\sbeta\mu\nu}\bigr)$ is the GB invariant that gives a
total derivative in four dimensions.
\par
We are interested in the variation over the metric
$\bg^{\alpha\sbeta}$, which after some manipulation it
gives\foot{See Ref.~[\fard] for a few useful equations.}\
$$
G^{(\rm SGB)}_{\alpha\sbeta}={1 \over 2}\kappa^2\,
T_{\alpha\sbeta}\ ,\eqno\eq
 $$
where we have arranged the result as $G^{(\rm
SGB)}_{\alpha\sbeta}=G^{(1)}_{\alpha\sbeta}+G^{(\phi)}_{\alpha\sbeta}
+G^{(0)}_{\alpha\sbeta}+f(\phi)G^{(2)}_{\alpha\sbeta}
+G^{\bigl(f(\phi)\bigr)}_{\alpha\sbeta}\equiv R^{(\rm
SGB)}_{\alpha\sbeta}-{1\over 2}\bg_{\alpha\sbeta}\, R^{(\rm
SGB)}$, and where also details of the corresponding
Euler--Lagrange expressions are $G^{(1)}_{\alpha\sbeta}\equiv
G_{\alpha\sbeta}=R_{\alpha\sbeta}-{1\over 2}\bg_{\alpha\sbeta}\,
R$,
$$
G^{(\phi)}_{\alpha\sbeta}\equiv R^{(\phi)}_{\alpha\sbeta}-{1\over
2}\bg_{\alpha\sbeta}\, R^{(\phi)}\equiv\left(-{\gamma\over
2}\partial_{\alpha}\phi\,\partial^{\sbeta}\phi\right)-{1\over
2}\bg_{\alpha\sbeta}\left(-{\gamma\over
2}\partial_{\mu}\phi\,\partial^{\mu}\phi\right)\ ,
 $$
$$
G^{(0)}_{\alpha\sbeta}\equiv R^{(0)}_{\alpha\sbeta}-{1\over
2}\bg_{\alpha\sbeta}\, R^{(0)}\equiv\left[{-V(\phi)\over
D-2}\bg_{\alpha\sbeta}\,\right]-{1\over
2}\bg_{\alpha\sbeta}\left[{-V(\phi)\over D-2}\right]={1\over
2}V(\phi)\bg_{\alpha\sbeta}\ ,
 $$
see Ref.~[\fare] for details,
$$
\eqalign{G^{(2)}_{\alpha\sbeta}
 &\equiv R^{(2)}_{\alpha\sbeta}-{1\over 2}\bg_{\alpha\sbeta}\,
  R^{(2)}\cr
 &\equiv\biggl[2R\,
  R_{\alpha\sbeta}-4\Bigl(R_{\alpha\mu}\, R_{\sbeta}{}^{\mu}
  +R_{\alpha\mu\sbeta\nu}\, R^{\mu\nu}\Bigr)+2R_{\alpha\rho\mu\nu}\,
  R_{\sbeta}{}^{\rho\mu\nu}\biggr]-{1\over 2}\bg_{\alpha\sbeta}\,
  \kappa^2 L^{(2)}\cr }
 $$
and
$$
\eqalign{G^{\bigl(f(\phi)\bigr)}_{\alpha\sbeta}
 &\equiv R^{\bigl(f(\phi)\bigr)}_{\alpha\sbeta}-{1\over
  2}\bg_{\alpha\sbeta}\, R^{\bigl(f(\phi)\bigr)}\cr
 &\equiv\biggl[-2R\, f(\phi)_{;\,\alpha\sbeta}+4R_{\alpha\mu}\,
  f(\phi)_{;\,\sbeta}{}^{\mu}+4R_{\sbeta\mu}\, f(\phi)_{;\,\alpha}{}^{\mu}
  -4R_{\alpha\sbeta}\, f(\phi)_{;\,\mu}{}^{\mu}\cr
 &\quad +4R_{\alpha\mu\sbeta\nu}\, f(\phi)^{;\,\mu\nu}-{1\over
  2}\bg_{\alpha\sbeta}\, a\, R f(\phi)_{;\,\mu}{}^{\mu}-{1\over
  2}\bg_{\alpha\sbeta}\, b\, R^{\mu\nu}\, f(\phi)_{;\,\mu\nu}\biggr]\cr
 &\quad-{1\over
  2}\bg_{\alpha\sbeta}\,\biggl[\bigl(-4-a\bigr)R f(\phi)_{;\,\mu}{}^{\mu}
  +\bigl(8-b\bigr)R^{\mu\nu}\, f(\phi)_{;\,\mu\nu}\biggr]\ ,\cr}
 $$
where $a$ and $b$ are arbitrary constants, $(;\mu\nu)$\ means
$(;\mu\, ;\nu)$ and so on.
\par
All the Ricci--like tensors and the Ricci--like scalars are
defined in a way to satisfy the necessarily corresponding trace
relations, namely ${\sl Trace}\, R^{(\rm
SGB)}_{\alpha\sbeta}=R^{(\rm SGB)}$,\ ${\sl Trace}\,
R_{\alpha\sbeta}=R$,\ ${\sl Trace}\,
R^{(\phi)}_{\alpha\sbeta}=R^{(\phi)}$,\ ${\sl Trace}\,
R^{(0)}_{\alpha\sbeta}=R^{(0)}$,\ ${\sl Trace}\,
R^{(2)}_{\alpha\sbeta}=R^{(2)}$\ and finally ${\sl Trace}\,
R^{\bigl(f(\phi)\bigr)}_{\alpha\sbeta}=R^{\bigl(f(\phi)\bigr)}$\
when the constants $a$ and $b$ are:
$$
a=-{2\over D-2}\quad\qquad {\rm and}\quad\qquad b={8\over
D-2}\eqno\eq
 $$
for $D\neq 2$. That is, the latter trace relation depends on the
dimension of space--time, as one may have expected due to the GB
term, with which a similar deduction has been arisen in
Ref.~[\fare].
\par
Therefore, we have been able to develop generalized Einstein
tensor technique for the SGB gravity via the Trace operator.
\medskip
\par
\noindent
 {\bf Acknowledgment}\nextline
\noindent

The author is grateful to Prof. John M. Charap and the Physics
Department of Queen Mary \& Westfield College University of London
where some part of this work was carried out.
%%%%%%%%
\tenpoint
\refout
%%%%%%%%
\bye